\documentclass[twocolumn,secnumarabic,amssymb,nobibnotes,aps,prx,showpacs,superscriptaddress,longbibliography]{revtex4-2}
\usepackage[english]{babel}
\usepackage{graphicx}
\usepackage{lmodern}
\usepackage{amsmath}
\usepackage{amsfonts}
\usepackage[usenames,dvipsnames]{color}
\usepackage{hyperref}
\hypersetup{colorlinks}
\hypersetup{citecolor=blue}
\usepackage{blindtext}
\usepackage[utf8]{inputenc}
\usepackage{color}
\usepackage{mathtools}
\usepackage[normalem]{ulem}
\usepackage{bm}
\usepackage[babel]{microtype}
\usepackage{units}
\usepackage[title]{appendix}
\usepackage[version=4,arrows=pgf-filled]{mhchem}
\usepackage{textcomp}
\usepackage{comment}

\makeatletter
\newcommand{\customlabel}[2]{%
   \protected@write \@auxout {}{\string \newlabel {#1}{{#2}{\thepage}{#2}{#1}{}} }%
   \hypertarget{#1}{}
}

\def\*#1{\mathbf{#1}}
\def\!#1{\mathbf{\hat#1}}

\newcommand{\JE}[1]{\textcolor{blue}{[JE: #1]}}

\newcommand{\dd}{\mathrm{d}}
\newcommand{\pKa}{\mathrm{p}K_\mathrm{a}}

\begin{document}
\title{Packaged DNA genome sets the long-range electrostatic anisotropy of a virus}
\author{Jeffrey C. Everts}
\altaffiliation{These authors contributed equally to this work. J.~C.\ Everts: \href{{jeffrey.everts@fuw.edu.pl}}{{jeffrey.everts@fuw.edu.pl}}, A.\ Božič: \href{{anze.bozic@ijs.si}}{{anze.bozic@ijs.si}}}
    \affiliation{Institute of Theoretical Physics, Faculty of Physics, University of Warsaw, Pasteura 5, 02-093 Warsaw, Poland}
        \affiliation{Institute of Physical Chemistry, Polish Academy of Sciences, 01-224 Warsaw, Poland}
     
\author{Anže Božič}
\altaffiliation{These authors contributed equally to this work. J.~C.\ Everts: \href{{jeffrey.everts@fuw.edu.pl}}{{jeffrey.everts@fuw.edu.pl}}, A.\ Božič: \href{{anze.bozic@ijs.si}}{{anze.bozic@ijs.si}}}
\affiliation{Department of Theoretical Physics, Jožef Stefan Institute, SI-1000 Ljubljana, Slovenia}

\author{Rudolf Podgornik}
\email{Deceased 28/12/2024}
\affiliation{School of Physical Sciences, University of Chinese Academy of Sciences, Beijing 100049, China}
\affiliation{Kavli Institute for Theoretical Sciences, University of Chinese Academy of Sciences, Beijing 100049, China}
\affiliation{Wenzhou Institute, University of Chinese Academy of Sciences, Wenzhou 325001, China}
\date{\today}

\begin{abstract}
Viruses are among the most highly charged objects in biology, and electrostatic interactions permeate nearly every stage of their life cycle. At the same time, an empty icosahedral capsid is almost isotropic as far as the surrounding electrolyte is concerned---its charge distribution is anisotropic only at high multipole order, and electrostatic screening removes such anisotropy long before it can be probed at a distance. Here, we show that the packaged genome of a filled virus imparts a long-range electrostatic anisotropy that extends throughout the surrounding medium, which is not present for empty capsids. We support these findings with a novel Poisson--Boltzmann continuum theory where the orientational order of double-stranded DNA enters through an anisotropic dielectric tensor and an inhomogeneous volume-charge distribution, which is coupled to a charge-regulating shell containing ionizable amino acid residues of the capsid proteins. The resulting electrostatic potential profiles show that DNA packed in an inverse-spool geometry imprints a highly inhomogeneous potential distribution on the outer capsid surface, whose angular structure is a direct result of the DNA-free axial void. At high salinity, the multipole spectrum of this distribution persists at large distances, whereas at low salt concentration, higher-order multipoles decay more rapidly away from the surface. The surviving quadrupole renders virus--wall and virus--virus interactions orientation dependent at the $k_\mathrm{B}T$ level, with a preferred orientation controlled by the sign of the external charge and the pH. Using phage $\lambda$ as a model system and checking robustness of our results on two other phages, we find the same behavior across a range of pH and ionic strengths. This identifies the packaged genome, rather than the symmetry of the capsid, as the origin of the long-range electrostatic anisotropy of a filled virus.
\end{abstract}

\maketitle
\begingroup
\renewcommand{\thefootnote}{*}
\footnotetext{These authors contributed equally to this work.}
\endgroup

\section{Introduction}

In their simplest form, viruses consist of a genome---most often double-stranded deoxyribonucleic acid (dsDNA) or single-stranded ribonucleic acid (ssRNA)---enclosed by a proteinaceous shell termed the capsid~\cite{carter2007virology}. Under typical environmental conditions both the capsid proteins and the genome acquire charge, making viruses among the most highly charged objects in biology, carrying tens of thousands of elementary charges~\cite{bozic2012simple,phillips2012physical}. Electrostatic interactions therefore permeate the viral life cycle, governing capsid assembly and genome packaging, interactions with cellular membranes and their receptors, as well as virus stability, morphological transitions, and disassembly~\cite{karlin1988charge,siber2012energies,carrivain2012electrostatics,cardoso2023physical,garmann2014role,javidpour2021electrostatic,bruinsma2021physics,panahandeh2022virus,heffron2021virus,garmann2026packaging,duran2021controlling}. However, how much of the effective electrostatic interaction of a virus with its environment is caused by the encapsidated genome and, in particular, its packing geometry remains an open question.

\begin{figure}[t]
\centering
\includegraphics[width=0.48\textwidth]{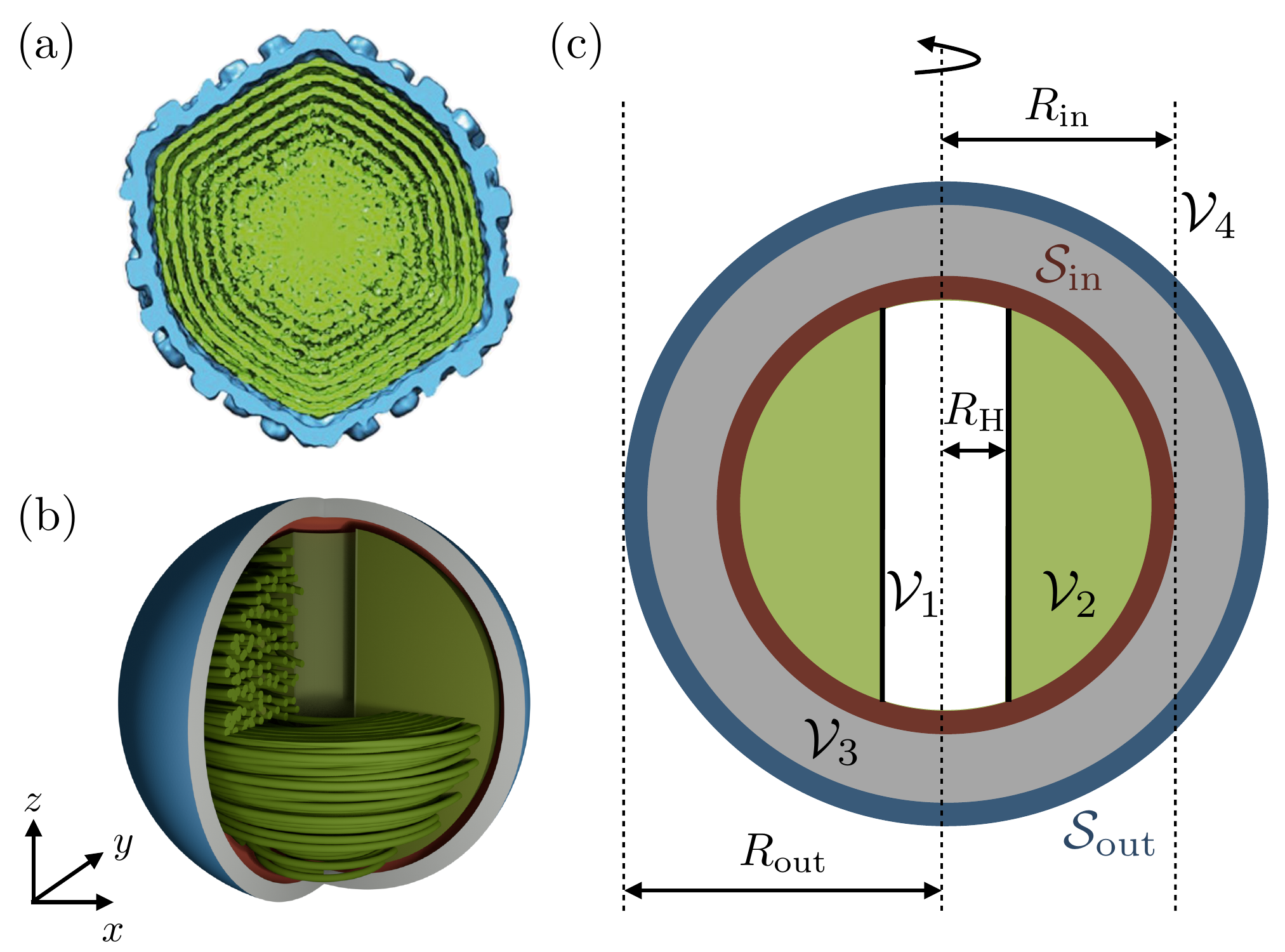}
\caption{{(a)} Cutaway view of phage $\lambda$ cryo-EM reconstruction with completely packaged DNA. {(b)} Three-dimensional cut-out sketch of a generic bacteriophage with nematically ordered DNA packaged inside a spherical capsid in an inverse spool with a central DNA-free void. {(c)} Cross-section of the model system with annotated geometric quantities defined in the main text. The model is composed of an axially symmetric DNA spool with an excluded central cylindrical region, surrounded by a capsid modelled as a double spherical shell with an inner and outer radius. The entire virus is surrounded by an external solvent. Panel (a) adapted from Ref.~\cite{lander2013dna} under CC BY-NC 3.0 licence.
}
\label{fig:1}
\end{figure}

This question is particularly relevant for dsDNA bacteriophages, which constitute the majority of known viruses and are the most abundant of all living organisms~\cite{hendrix2002bacteriophages,hatfull2025all}. Their genomes are long ($\gtrsim 10^4$ base pairs and beyond~\cite{hatfull2008bacteriophage,ryu2016molecular}) and must be driven into spherical capsids by molecular motors against the elastic and especially electrostatic forces that resist DNA compaction~\cite{siber2012energies,prevo2024dna}. At full packing the encapsidated DNA of many phages approaches near-crystalline densities and develops local nematic order~\cite{podgornik2016dna,zandi2020virus,coshic2024structure}. An example of a cryo-EM reconstruction of phage $\lambda$~\cite{lander2013dna} with visible genome ordering is shown in Fig.~\ref{fig:1}(a). Although no single conformation prevails---the packaged genome samples a family of configurations set by system-specific details~\cite{coshic2024structure,leforestier2013polymorphism,liu2014solid,petrov2007conformation,farrell2024spool}---these structures share global properties such as pressure, energy, and average nematic order parameter~\cite{coshic2024structure}. For analytical work outside simulations, the ordered packing is usually captured by some variant of the inverse-spool model (sketched in Fig.~\ref{fig:1}(b)), in which the DNA is arranged as coaxial hoops stacked along a common axis and locally in a nematic phase~\cite{ubbink1995polymer,purohit2003mechanics,evilevitch2003osmotic,purohit2005forces,marenduzzo2010biopolymer,shin2011filling,liu2021ion,tzlil2003forces,jiang2006structure}.

The presence of a tightly packaged DNA genome measurably alters the virus, affecting capsid structure, stiffness, and stability~\cite{van2018effect,carrasco2006dna,nam2023distinguishing} as well as, more surprisingly, the hydrophobicity and charge sensed at the outer capsid surface~\cite{hernando2015quantitative,heldt2023empty,caniglia2022probing}. While the charge measured outside an empty and a DNA-filled capsid differs, direct transfer of charge from the genome across the proteinaceous shell is energetically implausible and has been deemed unlikely~\cite{hernando2015quantitative,heldt2023empty}. Part of the explanation is straightforward and does not depend on how the genome is arranged---the interior charge raises the magnitude of the electrostatic potential throughout the particle, and a capsid whose ionizable residues are in chemical equilibrium with the solution responds by shifting their protonation state~\cite{nap2014role,kusters2015role}. What is unclear, however, is how this impacts the anisotropic footprint of the empty and filled capsids, which has a direct influence on their electrostatic interaction with the environment.

An empty icosahedral capsid is highly charged, and the distribution of that charge over its surface is far from uniform~\cite{bozic2012simple}. To its surroundings, however, it appears almost isotropic---icosahedral symmetry admits no anisotropic multipole below $l=6$, and in an electrolyte the higher the multipole, the more rapidly it is screened away~\cite{bozic2013symmetry}. Whatever orientational information the capsid surface carries is therefore erased within a fraction of the Debye screening length. This is the same mechanism that makes the effective interactions of patchy colloids and proteins so sensitive to which low-order multipoles their charge patterns contain~\cite{Bianchi2011,deGraaf2012,Gnidovec:2025,pineda2025}: at any appreciable separation, an electrolyte transmits only the lowest few.
 
This exposes the challenge we address here: a packaged genome is not constrained by the symmetry of the capsid that contains it. If the interior charge is arranged with a lower symmetry than icosahedral, the electrostatic anisotropy of a virus seen by its environment may extend beyond the one imposed by its capsid. The inverse-spool ordering of dsDNA breaks the interior symmetry down to a cylindrical one, whose leading anisotropy is quadrupolar ($l=2$) and therefore screened far more weakly than the icosahedral one. Whether such an interior anisotropy survives passage through a low-permittivity, charge-regulating protein shell is not obvious a priori, and it is what we determine here. To do so, we develop a minimal continuum model whose central ingredient is an electrostatic description of a nematically ordered genome, in which the orientational order enters explicitly through both an anisotropic dielectric response and the symmetry of the packaged charge; to our knowledge such a formulation has not been developed before. Anisotropic screening in an orientationally ordered medium has been studied for colloids suspended in nematic solvents~\cite{everts2021anisotropic}, but not, as here, for an ordered medium confined inside a dielectric shell. We couple the genome to a charge-regulating capsid~\cite{ninham1971electrostatic,markovich2016charge,Trefalt2016,Lund2013,Avni2019,bozic2026closed} whose ionizable amino acid residues---their types and their partition between the inner and outer surfaces---are taken from the structure of phage $\lambda$~\cite{lander2013dna}. With this model we establish under what conditions the interior (genome) charge and its anisotropy reach the exterior, characterize the resulting imprint on the surface charge and potential, and trace its consequences for the orientation-dependent interaction of a virus with a charged wall and with another virus.

The paper is outlined as follows. Section~\ref{sec:model} introduces a minimal virus geometry, and Sec.~\ref{sec:freeen} the electrostatic theory of an ordered dsDNA genome inside a charge-regulating shell. Section~\ref{sec:singlecapsid} then establishes that the genome \emph{does} leave an imprint on the outer surface, and that the imprint has a simple geometric origin. Section~\ref{sec:multicapsids} next demonstrates that the part of the imprint that survives screening is large enough to make the interaction of a virus both with a charged wall and with another virus depend on its orientation. Section~\ref{sec:exp} describes the measurements which could detect these effects, and Sec.~\ref{sec:dis} discusses the limitations of the work as well as its broader implications.

\section{Virus model}
\label{sec:model}

To study how the encapsidated DNA gives rise to anisotropic electrostatic effects, we consider a spool-like DNA confined in a spherical capsid~\cite{ubbink1995polymer,purohit2003mechanics}, with the geometry shown in Fig.~\ref{fig:1}(c). The spool axis is taken to be oriented in the $z$ direction. Our model consists of four regions $\mathcal{V}_{i}$ ($i=1,...,4)$ modelled as dielectric continua with sharp boundaries. The DNA-filled capsid  consists of a central cylindrical region $\mathcal{V}_1$ with radius $R_\mathrm{H}$, filled with solvent and devoid of DNA; region $\mathcal{V}_2$ filled with a composite composed of DNA and water; and a confining charged dielectric spherical shell $\mathcal{V}_3$ consisting of capsid proteins, characterized by an inner and outer radius $R_\mathrm{in}$ and $R_\mathrm{out}$. The exterior of the capsid (region $\mathcal{V}_4$) is filled with solvent. The two relevant boundaries of the system are between the inner capsid and the DNA spool ($\mathcal{S}_\mathrm{in}$) and between the outer capsid and the external solvent ($\mathcal{S}_\mathrm{out}$). With the exception of $\mathcal{V}_3$, ions are present in the entire system volume and are treated in a grand-canonical fashion. For simplicity, an equal number of monovalent cation and anion types is assumed, where the ionic strength $I= {\kappa_\mathrm{D}^2}/(8 \pi \ell_\mathrm{B})$ is a measure of the total bulk ion density. Here, $\kappa^{-1}$ is the Debye length and $\ell_\mathrm{B}=e^2/(4\pi\varepsilon_0\varepsilon k_\mathrm{B}T)=0.74$~nm is the Bjerrum length of water, with $e$ the elementary charge, $\varepsilon_0$ the vacuum permittivity, $\varepsilon=80$ the relative permittivity of water, $k_\mathrm{B}$ Boltzmann constant, and $T=298$~K (room) temperature.

Our model is deliberately minimal in one crucial aspect: by taking the capsid to be spherical we remove every source of exterior anisotropy other than the genome itself, so that any angular structure we find outside the particle can be attributed unambiguously to what is packaged inside. We fix the geometry of the system and take the inner and outer radius of the capsid shell to be $R_\mathrm{in}=28.8$~nm and $R_\mathrm{out}=32.1$~nm, respectively---values characteristic of a large number of bacteriophages~\cite{bozic2012simple}, and specifically of phage $\lambda$, which is used to derive virus-related parameters of our model. The capsid structure deposited in VIPERdb~\cite{montiel2021viperdb} (entry 7vii) forms the basis for most of the capsid parameters, as we show in the Supplemental Material (SM)~\cite{SM}. The radius of the cylindrical void $R_\mathrm{H}$ effectively corresponds to the amount of packaged DNA in base pairs (bp) via
\begin{equation}
 N_\mathrm{DNA}=\frac{4v}{3}\frac{\left(R_\mathrm{in}^2-R_\mathrm{H}^2\right)^{3/2}}{R_\mathrm{bp}^2 L_\mathrm{bp}},
\end{equation}
where the elementary DNA base pair unit is modelled as a cylinder with length $L_\mathrm{bp} = 0.34~{\rm nm}$~\cite{phillips2012physical} and radius $R_\mathrm{bp}=1$ nm~\cite{arnott1972optimised}. To estimate $v$, we note that a composite medium of oriented anisotropic cylinders with hexagonal close packing has a volume fraction $v_\mathrm{HCP} = 0.91$, and we therefore choose a slightly lower value of $v = 0.8$. This corresponds to a hexagonal interaxial spacing of $2.13$~nm, at the dense end of the $2.4$--$2.8$~nm range reported for fully packaged phages~\cite{earnshaw1977dna,evilevitch2003osmotic}. As will become clear in Sec.~\ref{sec:singlecapsid}, the value of $v$ enters the amplitude of the genome imprint but not its angular structure, which is controlled by $R_\mathrm{H}$ and the capsid geometry. We choose $R_\mathrm{H}=14.5$~nm, which roughly yields $N_\mathrm{DNA}=48.4$~kbp of DNA contained in the capsid---the length of the phage $\lambda$ genome and in general a typical genome length encountered in bacteriophages of this size~\cite{hatfull2008bacteriophage,luque2020missing}.

Since the virus model we use is fairly complex---despite the approximations involved---we keep most of the system parameters fixed and vary those that show the largest influence on the observed electrostatic anisotropy. A summary of all the \emph{fixed} system parameters for this study is listed in Table~S1 of the SM \cite{SM}.

\section{Electrostatic description of a single virus}

\label{sec:freeen}
The electrostatic free energy $F_\mathrm{c}$ of a single capsid containing a given (static) DNA configuration can be expressed solely in terms of the (fluctuating) electrostatic potential $\psi({\bf r})$ via the functional integral $e^{-\beta F_\mathrm{c}}\propto \int\mathcal{D}\psi\, e^{-\beta\mathcal{F}_\mathrm{c}[\psi]}$, where $\beta^{-1}=k_\mathrm{B}T$ is the thermal energy. Here, we used the fact that the free energy functional $\mathcal{F}_\mathrm{c}[\psi]$ of a single capsid in an unbounded electrolyte does not depend on its center-of-mass position nor the orientation of its spool axis. We decompose $\mathcal{F}_\mathrm{c}[\psi]$ according to the defined regions and boundaries (Sec.~\ref{sec:model} and Fig.~\ref{fig:1}(c)):
\begin{equation}
\mathcal{F}_\mathrm{c}[\psi]=\sum_{i\in\{\textrm{in,out}\}}\mathcal{F}_{\mathrm{S},i}[\psi]+\sum_{i=1}^4\mathcal{F}_{\mathrm{V},i}[\psi]. \label{eq:freeenergy}
\end{equation}
The surface free energies (first term) set the boundary conditions for all interfaces and will be specified later. First, we focus on the bulk (volume) contributions $\mathcal{F}_{\mathrm{V},i}[\psi]$.

\subsection{Bulk contributions}

The regions $\mathcal{V}_1$ and $\mathcal{V}_4$ are filled with water and free ions, contributing to the free energy of a single capsid
\begin{align}
\mathcal{F}_{\mathrm{V},i}[\psi]=&
- \int_{\mathcal{V}_{i}} \dd V\, \Bigg[   \frac{1}{2}\varepsilon_0  \varepsilon |\nabla \psi({\bf r})|^2\nonumber\\
&+ 2k_\mathrm{B}TI\cosh{\beta e \psi({\bf r}) }\Bigg], \quad i=1,4.
\label{eq:F1F4}
\end{align} 
On the other hand, the protein region $\mathcal{V}_3$ does not contain any ions and is described by
\begin{align}
\mathcal{F}_{\mathrm{V},3}[\psi]&=
- \frac{1}{2} \varepsilon_0  \varepsilon_\mathrm{p}\int_{\mathcal{V}_3} \dd V\,   |\nabla \psi({\bf r})|^2,
\label{eq:F3}
\end{align} 
with $\varepsilon_\mathrm{p}=4$ the relative permittivity of the capsid (protein) interior~\cite{amin2020variations}.

Equations~\eqref{eq:F1F4} and~\eqref{eq:F3}, describing regions $\mathcal{V}_1$, $\mathcal{V}_3$, and $\mathcal{V}_4$, are standard for Poisson--Boltzmann (PB) type theories. However, to describe the DNA-filled region $\mathcal{V}_2$, we propose {\em a novel method} that will allow us to investigate the effects of (local) orientational ordering on the electrostatics of the system. Specifically, we assume that the DNA is nematically ordered in an inverse spool geometry with order parameter $\bm{\mathsf{Q}}$ and charge density $\rho_\mathrm{DNA}({\bf r})$:
\begin{equation}
\begin{split}
\mathcal{F}_{\mathrm{V},2}[\psi]= &
- \int_{\mathcal{V}_2} \dd V\, \Bigg[   \frac{1}{2}\varepsilon_0  \nabla \psi({\bf r})\cdot\boldsymbol{\varepsilon}(\bm{\mathsf{Q}})\cdot\nabla\psi({\bf r})\\
 &-\rho_\mathrm{DNA}({\bf r})\psi({\bf r})+ 2k_\mathrm{B}TI\cosh{\beta e \psi({\bf r}) }\Bigg],
 \end{split}
\label{eq:F2}
\end{equation} 
with $\boldsymbol{\varepsilon}(\bm{\mathsf{Q}})$ the dielectric tensor of the anisotropic DNA--water composite. 
The dielectric tensor is parametrized as
\begin{equation}
\boldsymbol{\varepsilon}(\bm{\mathsf{Q}})=\frac{1}{3}(2\varepsilon_\perp+\varepsilon_\parallel)\,\bm{\mathsf{I}}+\frac{2}{3}(\varepsilon_\parallel-\varepsilon_\perp)\,\bm{\mathsf{Q}},
\end{equation}
which is a common description for nematics \cite{deGennes}.
We assume a fully developed nematic phase in the condensed inverse spool with nematic order parameter $\bm{\mathsf{Q}}=(3/2)[\hat{\bf n}\hat{\bf n}-(1/3)\bm{\mathsf{I}}]$ and director pointing in the azimuthal direction, $\hat{\bf n}=\hat{\boldsymbol{\varphi}}$ (using cylindrical coordinates $(\varrho,\varphi,z)$).

The DNA phase is treated as a dielectric composite of rods (DNA) and an isotropic matrix (water). The Rayleigh formula for the anisotropic dielectric constant of such a composite is~\cite{hopkins2015disentangling}
\begin{equation}
{\varepsilon_{\parallel}}=\varepsilon\left(1+v\Delta_{\parallel}\right), \quad {\varepsilon_{\perp}}=\varepsilon\left(1+\frac{2v\Delta_{\perp}}{1-v\Delta_{\perp}}\right),
\end{equation}
 with $v$ the volume fraction of DNA in the composite and 
 \begin{equation}
\Delta_{\perp}=\frac{{\varepsilon^{c}}_{\perp}-\varepsilon}{{\varepsilon^{c}}_{\perp}+\varepsilon}\quad\mathrm{and}\quad
\Delta_{\parallel}=\frac{{\varepsilon^{c}}_{\parallel}-\varepsilon}{\varepsilon},
\end{equation}
where ${\varepsilon^{c}}_{\perp}$  and ${\varepsilon^{c}}_{\parallel}$ are the transverse and longitudinal dielectric response functions of the cylinder material. These can be taken to be the same, ${\varepsilon^{c}}_{\perp} \simeq {\varepsilon^{c}}_{\parallel} = 2$, typical for non-conducting polymeric regions.

We note here that the tensorial nature of the dielectric tensor is irrelevant in the cylindrically-symmetric single-capsid geometry. In cylindrical coordinates, the electrostatic potential $\psi({\bf r})$ does not depend on $\varphi$, and therefore only $\varepsilon_{\varrho\varrho}$ and $\varepsilon_{zz}$ contribute to the free energy---and both these quantities are equal to $\varepsilon_\perp$. In contrast, in two-capsid geometries and in the presence of a charged wall, there are also contributions from $\varepsilon_\parallel$. This implies that in every axisymmetric configuration---the single isolated capsid, and the capsid--wall geometry with the spool axis along the surface normal---the nematic order enters the electrostatics through the in-plane permittivity $\varepsilon_\perp$ of the composite and through the architecture that the ordering imposes, in particular the DNA-free axial void. The tensorial character of the response becomes operative only once the axial symmetry is broken, as is the case for a tilted spool near a wall or for a pair of capsids. Both aspects originate in the same orientational order but act in different geometries.

The charge density of DNA is negative due to the presence of phosphate groups and depends on the (local) nematic order, and we can expand it as
\begin{equation}
\rho_\mathrm{DNA}({\bf r})\psi({\bf r})=n_\mathrm{bp}v\left(-e+t\bm{\mathsf{Q}}:\nabla\nabla+\ldots\,\right)\psi({\bf r}),  \label{eq:localmulti}
\end{equation}
which should be seen as a multipole expansion followed by a local approximation, where we only consider terms up to the quadrupole. Here, $n_\mathrm{bp}$ is the number density of phosphate groups in the condensed inverse spool and $t\bm{\mathsf{Q}}$ is the quadrupole tensor of DNA assumed to be proportional to the nematic order parameter, where $t$ is the electrical quadrupole moment of the DNA. Furthermore, since there are two charges per base pair, we estimate that  $n_\mathrm{bp} = 2/(\pi R_\mathrm{bp}^2 L_\mathrm{bp})$, with the $R_\mathrm{bp}$ and $L_\mathrm{bp}$ values from Sec. \ref{sec:model}.  

\subsection{Surface contributions: constant charge and charge regulation}

In this work, we consider both constant charge (CC) and charge regulation (CR) boundary conditions, which serve different purposes. Holding the surface charge fixed removes the capsid as a source of anisotropy, so that any imprint on the external electrostatic potential can only have come from the packaged genome. Allowing the surface charge to regulate restores the situation of a real capsid, whose ionizable residues are free to respond to the interior field in turn, and allows us to see not only the influence of the genome but also the response of the capsid.

In the CC case, the surface charge densities $\sigma_i$ on the inner and outer surfaces are fixed. The surface contributions $\mathcal{F}_{\mathrm{S},i}[\psi]$ for the inner and outer surface are determined by 
\begin{equation}
\mathcal{F}^\mathrm{CC}_{\mathrm{S},i}[\psi]=\oint_{\mathcal{S}_i} \dd S\, \sigma_{i}\psi({\bf r}), \quad i\in\{\mathrm{in,out}\}. \label{eq:FCC}
\end{equation}
We also assume that there are no (free) surface charges present on the boundary between $\mathcal{V}_1$ and $\mathcal{V}_2$. However, for real viruses the surface charge densities and surface potentials are not fixed quantities but depend sensitively on the thermodynamic state of the system, given for instance by pH and temperature, described through the CR mechanism~\cite{ninham1971electrostatic,markovich2016charge,podgornik2018}. Thus, the surface charge on the proteinaceous capsid stems from amino acids of type $\alpha$ having side chains with chargeable chemical groups $\ce{X}_\alpha$ that are either acidic or basic, and can therefore be deprotonated (becoming negatively charged) or protonated (becoming positively charged), respectively. 

The (partial) charge of a chargeable amino acid side chain is governed by a distinct acid dissociation constant $K_{\mathrm{a},\alpha}$, or equivalently $\mathrm{p}K_{\mathrm{a},\alpha}=-\log_{10}K_{\mathrm{a},\alpha}/c_0$~\cite{pahari2019pkad} with the reference concentration $c_0=1$ mol/L. Generally, surface $\pKa$ values are different from bulk $\pKa$ values \cite{Levin:2019}. However, for simplicity we take the bulk values that are independent of the specific surface where the amino acids are residing; see the SM~\cite{SM} for the chosen values. Amino acids with negatively charged side chains are aspartic acid (ASP), glutamic acid (GLU), and tyrosine (TYR), whereas positively charged amino acids are arginine (ARG), lysine (LYS), and histidine (HIS); while cysteine (CYS) can carry negative charge, it typically participates in disulfide bonds and we consider it uncharged~\cite{nap2014role}. Specifically, the charging reactions are
\begin{align}
-\ce{X}_\alpha\ce{H &<=>[K\sb{\mathrm{a},\alpha}] -X}_\alpha^- + \ce{H+}, \ \alpha=\mathrm{ASP, GLU, TYR}, \nonumber \\
-\ce{X}_\alpha\ce{H+ &<=>[K\sb{\mathrm{a},\alpha}] -X}_\alpha\ce{+ H+}, \ \ \alpha=\mathrm{ARG, LYS, HIS}. \label{eq:reactions}
\end{align}
While the DNA genome is in principle also charge-regulating, it is reasonable to assume a constant (volume-)charge in this case because the $\pKa$ value of a phosphate group is typically $\sim 1$~\cite{thaplyal2014experimental}.

The contributions of different ionizable amino acid moieties to the surface free energy for $i\in\{\mathrm{in,out}\}$ are
\begin{align}
&\mathcal{F}^\mathrm{CR}_{\mathrm{S},i}[\psi]=\sum_\alpha\int_{\mathcal{S}_i} \dd S\, \Bigg(z_\alpha e n_{i,\alpha} \psi({\bf r})- k_\mathrm{B}T\, n_{i,\alpha} \label{eq:freesurf2}\\
&\times \ln\Big\{ 1 + \mathrm{exp}\big[{z_\alpha \beta e\psi ({\bf r}) + z_\alpha \ln{10}(\mathrm{pH} - \mathrm{p}K_{\mathrm{a},\alpha})}\big]\Big\}\Bigg) \nonumber,
\end{align}
where $\alpha$ runs over all amino acids listed in Eq.~\eqref{eq:reactions}. We defined the reservoir pH by $\mathrm{pH}=-\log_{10} [\mathrm{H}^+]_\infty/c_0$, with $[\mathrm{H}^+]_\infty$ the bulk H$^+$ concentration. According to Eq.~\eqref{eq:reactions}, we have $z_\alpha=-1$ for amino acids with an acidic side chain $\alpha=$ ASP, GLU, TYR, and $z_\alpha=1$ for a basic side chain $\alpha=$ ARG, LYS, HIS. Note that the sign of this parameter and the form of Eq.~\eqref{eq:freesurf2} depends not only on the sign of the charge, but also on whether the chemical reaction leading to a net charge is dissociative or associative. Unlike the $\pKa$ values, the surface number densities of chargeable sites $n_{i,\alpha}$ for each amino acid $\alpha$ differ on the inner and outer surface $\mathcal{S}_\mathrm{in}$ and $\mathcal{S}_\mathrm{out}$.  As with other virus parameters, we model the distribution of chargeable amino acids based on the phage $\lambda$ structure, with a summary listed in Table~S1 of the SM~\cite{SM}.

\section{Electrostatic imprint of the genome on a single capsid}
\label{sec:singlecapsid}

Our first step is to establish whether the packaged genome can be probed outside the capsid at all, which we do through the analysis of the electrostatic potential. We approach this in two stages: first with the surface charge held fixed, which isolates the effect of the interior geometry, and then with a charge-regulating surface, which is the situation of a real capsid in solution.

To investigate the effects of the anisotropy in the genome region on the electrostatics outside the virus, we focus purely on the analysis of the model at the mean-field level (appropriate for monovalent ions \cite{Levin:2002}) determined by the saddle-point value of $\mathcal{F}_\mathrm{c}[\psi]$; see Eq.~\eqref{eq:freeenergy}. The relevant Euler-Lagrange (EL) equations are determined by
\begin{equation}
\left.\frac{\delta \mathcal{F}_\mathrm{c}[\psi]}{\delta\psi({\bf r})}\right|_{\psi({\bf r})=\psi_\mathrm{MF}({\bf r})}=0.
\end{equation}
This gives rise to modified PB equations defined in the regions in Fig.~\ref{fig:1}(c). The anisotropic nature of the genome region is encoded in the spatially inhomogeneous dielectric profile caused by the ordering geometry, the tensorial nature of dielectric properties due to the nematic ordering, and the inhomogeneous (volume) charge distribution of DNA. The surface contributions give rise to jump or continuity conditions of the dielectric displacement on the interfaces between the various regions; detailed equations are given in the SM~\cite{SM}. It is important to note that in the CR case the surface free energy (Eq.~\eqref{eq:freesurf2}) together with Gauss' law of electrostatics gives a surface charge distribution that is self-consistently defined in terms of the dimensionless electrostatic potential $\phi({\bf r})=\beta e\psi_\mathrm{MF}({\bf r})$, where
\begin{equation}
\sigma_i( {\bf r} )=  \sum_\alpha\frac{e n_{i,\alpha}z_\alpha}{1+\mathrm{exp}\big[{z_\alpha\phi({\bf r})+z_\alpha\ln 10\,(\mathrm{pH}-\mathrm{p}K_{\mathrm{a},{\alpha}})}\big]},
\label{BC-CR-1}
\end{equation}
for ${\bf r}\in\mathcal{S}_i$, with $i\in\{\mathrm{in,out}\}$. In the CC case this is a fixed, prescribed value.

The EL equations governing the system together with the appropriate boundary conditions are numerically solved in COMSOL Multiphysics\textregistered\  using finite-element methods. The non-trivial boundary conditions are accounted for by a suitable choice of fluxes within this software package; other numerical details can be found in the SM~\cite{SM}.
\begin{figure}[tp]
\centering
\includegraphics[width=0.975\columnwidth]{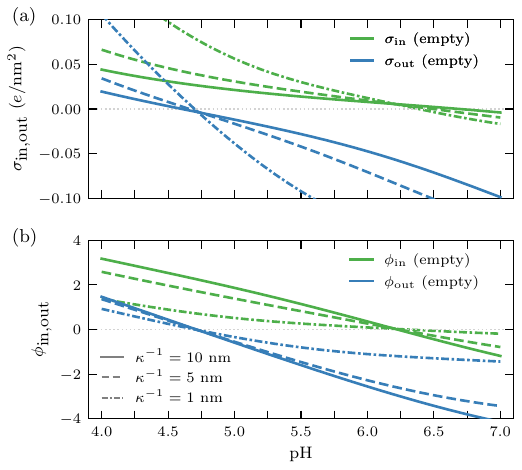}
\caption{Surface charge density and surface electrostatic potential of the $\lambda$ phage model with CR surface boundary conditions. (a) Inner and outer surface charge density and (b) inner and outer surface electrostatic potential of an empty capsid as a function of pH for three different values of the Debye screening length. Under CC boundary conditions, we set the same surface charge to the capsid that follows from panel (a) at a given pH and $\kappa^{-1}$.
}
\label{fig:2}
\end{figure}

\begin{figure*}
\centering
\includegraphics[width=0.9\textwidth]{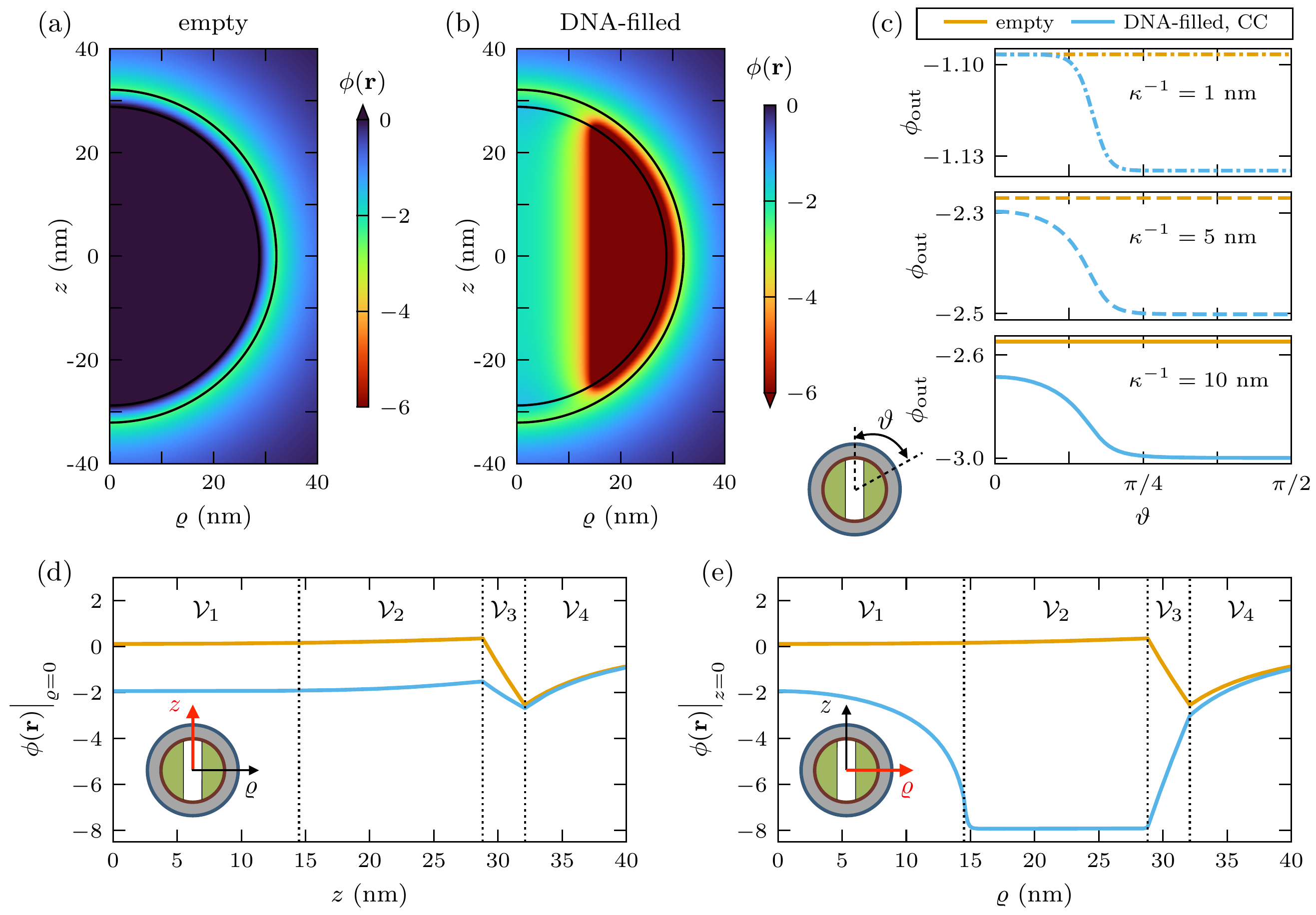}
\caption{Dimensionless electrostatic potential $\phi=\beta e\psi_\mathrm{MF}$ of the phage $\lambda$ model with CC surface boundary conditions (determined at $\mathrm{pH}=6$). The potential in the $(\varrho,z)$ plane (in cylindrical coordinates) is shown for {(a}) empty capsid (surface charge only) and {(b)} a capsid containing a charged DNA spool with only the monopole of the DNA charge retained ($t=0$). The two semi-circles in each panel denote the inner and outer capsid shells. {(c)} The potential on the outer capsid surface $\phi_\mathrm{out}$ as a function of the polar angle $\vartheta$ for both empty and filled capsids and for three values of the Debye screening length $\kappa^{-1}$. Panels (d) and (e) show the electrostatic potential along the $z$ and $\varrho$ axis, respectively, as shown in the insets. Dotted lines denote the boundaries between different regions of the capsid $\mathcal{V}_i$ (Fig.~\ref{fig:1}).
}
\label{fig:3}
\end{figure*}

\subsection{Anisotropy due to DNA geometry in constant surface charge conditions}
\label{ssec:CC}

First, we analyse an empty capsid to establish the charging behavior of our virus model. Since this geometry is isotropic, the electrostatic potential in the CC case is identical to the CR case if appropriate surface charges $\sigma_\mathrm{in}$ and $\sigma_\mathrm{out}$ are chosen. To make this specific, we determine in the CR case the surface charges $\sigma_\mathrm{in,out}$ and surface potentials $\phi_\mathrm{in,out}=\phi({\bf r})|_{{\bf r}\in\mathcal{S}_\mathrm{in,out}}$ for a range of thermodynamic states parametrized by the $\mathrm{pH}$ and $\kappa^{-1}$; see panels (a) and (b) of Fig.~\ref{fig:2}, respectively. The appropriate fixed surface charge in CC conditions at a given pH and $\kappa^{-1}$ is then determined from Fig.~\ref{fig:2}(a). With this choice, the CC and CR conditions are set to be identical in the empty capsid limit. Furthermore, note that our model allows for a charge reversal of both inner and outer capsid surface charge as a function of pH for a given $\kappa^{-1}$.

To highlight an example of an emergent electrostatic potential profile $\phi({\bf r})$, we fix a thermodynamic state by choosing $\mathrm{pH}=6$ and $\kappa^{-1}=10$ nm. For an empty capsid, the resulting isotropic $\phi({\bf r})$ is shown as a two-dimensional heatmap in Fig.~\ref{fig:3}(a); Fig.~\ref{fig:3}(c) demonstrates the absence of angular dependence of the surface potential on the polar angle $\vartheta$,  a consequence of the isotropic geometry and surface properties. Panels (d) and (e) of Fig.~\ref{fig:3} further show the electrostatic potential along two (identical) one-dimensional cuts in $\varrho$ and $z$ directions.

In the presence of a nematically ordered DNA genome, the situation is considerably different. Choosing the same $\sigma_\mathrm{in,out}$ as for the empty capsid case, we retain the homogeneous surface charge properties, but the region inside the capsid $\mathcal{V}_1\cup\mathcal{V}_2$ is anisotropic (cylindrically symmetric). Consequently, the resulting $\phi({\bf r})$ lacks spherical symmetry, especially inside the genome region, see Fig.~\ref{fig:3}(b). The lack of isotropy becomes further apparent in the one-dimensional cuts in Fig.~\ref{fig:3}(d),(e) and in the angular-dependent surface potential distribution in Fig.~\ref{fig:3}(c). The latter is a measure of the imprint of the genome which reaches outside the virus. The width of the surface potential distribution is $\sim8.1$~mV, which becomes smaller with decreasing $\kappa^{-1}$, see the dashed (middle panel) and dashed-dotted (top panel) blue lines in Fig.~\ref{fig:3}(c). From now on we focus only on outer surface properties, as these quantities typically determine electrophoretic properties~\cite{Teubner:1982} and effective interparticle interactions~\cite{Everts:2020}.

Already at fixed surface charge, the packaged genome produces an angular modulation of the exterior potential where an empty capsid produces none. From now on, we focus on the CR case in order to describe biologically relevant settings. The remainder of this section explores what determines the shape of the surface-potential anisotropy of a single capsid and how much of it survives once the capsid surface is free to respond.

\subsection{Long-range electrostatic imprint of packaged DNA through charge regulating capsid surface}
\label{ssec:CR}

\begin{figure}
\centering
\includegraphics{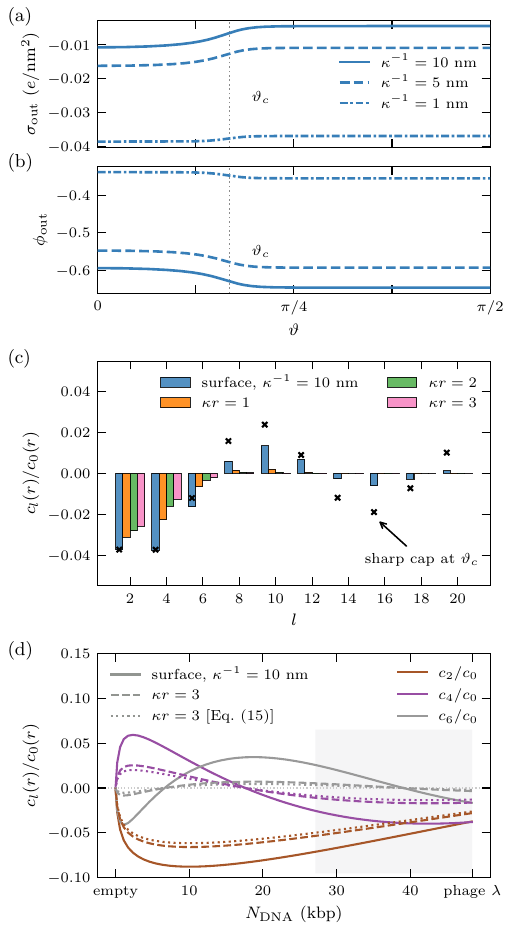}
\caption{Angular dependence of the (a) outer surface charge density and (b) outer surface potential of DNA-filled capsids with CR boundary conditions at $\mathrm{pH}=5$. (c) Distance-dependent ratio of the multipole expansion coefficients of the electrostatic potential relative to its monopole coefficient, $c_l(r)/c_0(r)$. Different colors correspond to different distances from the capsid surface $r$, starting with the outer surface potential and progressing in units of $\kappa r$. The symbols show the multipole expansion of the shape of a hard spherical cap with the angle $\vartheta_c$ (vertical dotted lines in panels (a) and (b)), corresponding to the angle at which the DNA-free void meets the inner capsid surface. (d) Distance-dependent ratio of the multipole coefficients $c_l(r)/c_0(r)$ as a function of the packaged DNA length. Shaded region shows the range where the DNA is expected to be in an ordered nematic phase ($L_\mathrm{packaged}\gtrsim27.1$~kbp). Dotted lines show the approximate decay of the multipole ratio at $\kappa r=3$ (Eq.~\eqref{eq:cl-rad}).
}
\label{fig:4}
\end{figure}

Allowing the capsid to regulate its charge changes the picture in a way that a fixed-charge surface cannot show: the genome now modulates both the surface potential and the surface charge itself, since the ionizable residues respond to the anisotropic interior field by locally shifting their protonation state (Fig.~\ref{fig:4}(a),(b)). The axisymmetric anisotropy can be quantified by expanding the surface fields in Legendre polynomials $P_l$; we do so for the surface potential as $\phi_\mathrm{out}(\vartheta)=\sum_l c_lP_l(\cos\vartheta)$, with expansion coefficients given by
\begin{equation}
    c_l = \frac{2l+1}{2} \int_{-1}^{1}\mathrm{d}u\,\phi_\mathrm{out}(u) P_l(u), \label{eq:multipoles}
\end{equation}
with $u=\cos\vartheta$. For a single capsid, all odd-$l$ coefficients vanish due to the symmetry of the spool.

The resulting spectrum (Fig.~\ref{fig:4}(c), blue bars) has a structure that is not what one would guess from the smooth profiles of Fig.~\ref{fig:4}(a),(b): the quadrupolar ($l=2$) and hexadecapolar ($l=4$) coefficients are of comparable size, and the higher coefficients alternate in sign. This is the signature of a two-level angular profile---one value within a polar cap with opening angle $\vartheta_\mathrm{c}$ located at either pole, and another value over the equatorial band. The spectrum of such a ``sharp'' cap is shown by the black symbols in Fig.~\ref{fig:4}(c) and qualitatively follows the data closely up to $l=20$. The cap angle is $\vartheta_\mathrm{c}\approx\arcsin(R_\mathrm{H}/R_\mathrm{in})$, the angle at which the DNA-free void meets the inner capsid surface. In other words, what the capsid presents to the outside is essentially the ``electrostatic shadow'' of the void inside it, blurred over an angular width comparable to the shell thickness seen from the center, given by $(R_\mathrm{out}-R_\mathrm{in})/R_\mathrm{out}$. The consequence is a clean separation of roles: the geometry of the packaged DNA fixes the angular shape of the imprint, while the electrostatic state of the system---pH, ionic strength, the charge of the DNA---enters only through its overall amplitude. That separation is what makes the anisotropy carry a footprint of the packaging rather than of the solution conditions, to which we return in Sec.~\ref{sec:dis}.

Equation~\eqref{eq:multipoles} can be evaluated on any spherical surface at fixed $r=|\mathbf r|$, which lets us follow the imprint at a distance from the capsid surface. Doing so (Fig.~\ref{fig:4}(c), colored bars) shows that the different multipoles do not simply fade together: the high-order coefficients are attenuated far more rapidly than the low-order ones. Screening thus acts as a filter on the angular pattern, and the pattern that reaches the surrounding medium is not the one that is imprinted on the capsid surface. While the lowest multipoles, $l=2$ and $l=4$, are comparable at the capsid surface, the quadrupole dominates a few screening lengths away.

Within linear screening this filtering of different multipoles has a closed form~\cite{bozic2013symmetry,bozic2026closed}. Each multipole propagates as $c_l(r)\propto k_l(\kappa r)$, with $k_l$ the modified spherical Bessel function of the third kind, so that
\begin{equation}
\frac{c_l(r)}{c_0(r)}=\frac{c_l(R_\mathrm{out})}{c_0(R_\mathrm{out})}\frac{\xi_l(\kappa r)}{\xi_l(\kappa R_\mathrm{out})},
\label{eq:cl-rad}
\end{equation}
where
\begin{equation}
\xi_l(x)=\sum_{n=0}^l\frac{(l+n)!}{n!(l-n)!}(2x)^{-n}.
\end{equation}
Since $\lim_{x\to\infty}\xi_l(x)=1$, the ratio $c_l(r)/c_0(r)$ does not decay to zero but saturates at $[c_l(R_\mathrm{out})/c_0(R_\mathrm{out})]\times[1/\xi_l(\kappa R_\mathrm{out})]$, implicating that beyond a few screening lengths all multipoles acquire the same radial dependence, and the angular pattern stops changing shape. Its final form is set by $\xi_l(\kappa R_\mathrm{out})$ alone. For phage $\lambda$ at $\kappa^{-1}=10$~nm, the ratio of the first two multipoles at the surface $c_4(R_\mathrm{out})/c_2(R_\mathrm{out})\approx1.015$ becomes $\approx0.18$ at large separation, which is the precise sense in which the quadrupole dominates the long-range anisotropy. The pink bars in Fig.~\ref{fig:4}(c) show that this linearized description is already accurate at distances a few screening lenghts away; it fails closer to the surface, where the potential is too large to linearize and nonlinear effects couple the modes.

Because $R_\mathrm{H}$ fixes the amount of packaged DNA, varying it let us examine how the anisotropic imprint develops as a capsid is filled with genome. Figure~\ref{fig:4}(d) shows the first three multipole ratios as the packaged length is increased from an empty capsid to the $48.4$~kbp of the phage $\lambda$ genome. Two features stand out. First, the ratios change substantially with filling, so the near-equality of the $l=2$ and $l=4$ components seen in Fig.~\ref{fig:4}(c) is a property of this particular void size rather than a general feature. The whole spectrum shifts as the angular size of the void changes, again showing that the geometry of the packaged DNA leaves its anisotropic footprint on the capsid surface. Second, the ordering of the multipoles at a distance does not change: at $\kappa r=3$ the higher components are suppressed at every filling (dashed lines), leaving the quadrupole dominant throughout. The dotted lines show the prediction of Eq.~\eqref{eq:cl-rad}, which reproduces the numerical result closely. We note that the DNA in our model would be expected to reach the assumed degree of order only above $\sim60\%$ packaging, so the physically meaningful range is $N_\mathrm{DNA}\gtrsim27.1$~kbp (shaded); below it, the curves should be read as showing the geometric structure of the model rather than the behavior of a partially filled phage.

These conclusions do not depend on our particular choice of the model system. In the SM~\cite{SM} we repeat the analysis over the pH range $4$--$6$, over screening lengths from $1$ to $10$~nm, and for two further phages, Sf6 and JBD30, whose capsid dimensions, amino acid compositions and genome lengths all differ from those of phage $\lambda$. In every case the surface profile retains the same two-level shape set by the void angle, described qualitatively by the same hard-cap spectrum; what changes is the amplitude. The screening length additionally controls how strongly the spectrum is filtered as one moves away from the capsid surface, in the manner made explicit by Eq.~\eqref{eq:cl-rad}: at $\kappa^{-1}=1$~nm the capsid is large compared with the screening length, the multipoles decay at nearly the same rate, and the pattern reaching the medium resembles the one on the surface much more closely than at $\kappa^{-1}=10$~nm. Salt therefore controls not only the range of the electrostatic field but also which part of its angular structure remains visible. The genome-induced lowering of the leading exterior anisotropy to quadrupolar order persists across all regimes we examined.

\subsection{Quadrupole component of DNA charge density}

Throughout this section we retained only the monopole of the DNA charge density in Eq.~\eqref{eq:localmulti}. To estimate the next-order contribution we note that a non-vanishing $t$ gives rise to a flexoelectric polarization field $\mathbf{P}=n_\mathrm{bp}vt\,\nabla\cdot\mathbf{Q}$. For an azimuthal director field the associated volume bound-charge density vanishes identically in $\mathcal V_2$, and the polarization instead produces a bound surface charge on the two interfaces at which the director field terminates: $\sigma_\mathrm{b}=3n_\mathrm{bp}vt/(2R_\mathrm{H})$ on the void boundary and $-3n_\mathrm{bp}vt/(2R_\mathrm{in})$ on the portion of $\mathcal S_\mathrm{in}$ in contact with DNA, carrying equal and opposite total charge. The second of these is uniform over exactly the equatorial band $|\cos\vartheta|<\cos\vartheta_\mathrm{c}$ that is bound by the cap geometry of the imprint of the DNA-free void. The quadrupole term can therefore rescale the amplitude of the imprint, but cannot alter its angular shape. Choosing $t$ such that $\sigma_\mathrm{b}$ lies between $\sim-0.1$ and $-1\,e/\mathrm{nm}^2$, we find the resulting change in the anisotropy to be roughly an order of magnitude smaller than the effects discussed above, and we therefore neglect it in what follows.

\section{Anisotropy in effective interactions}
\label{sec:multicapsids}

The quadrupole component that survives screening is small in absolute terms, and the question remain whether the effect is measurable in an experiment. We therefore turn to effective interactions, in two geometries that between them cover the situations in which a virus typically finds itself: beside an extended surface with a fixed charge, and beside another virus. In both, the quantity of interest is not the interaction strength but its dependence on the orientation of the spool axis, since that is the part of the interaction that an isotropic particle does not have. The free energy functionals take the same form as in Sec.~\ref{sec:freeen}, but their minima now depend on the configuration of the particles.

\begin{figure*}
\centering
\includegraphics[width=0.88\textwidth]{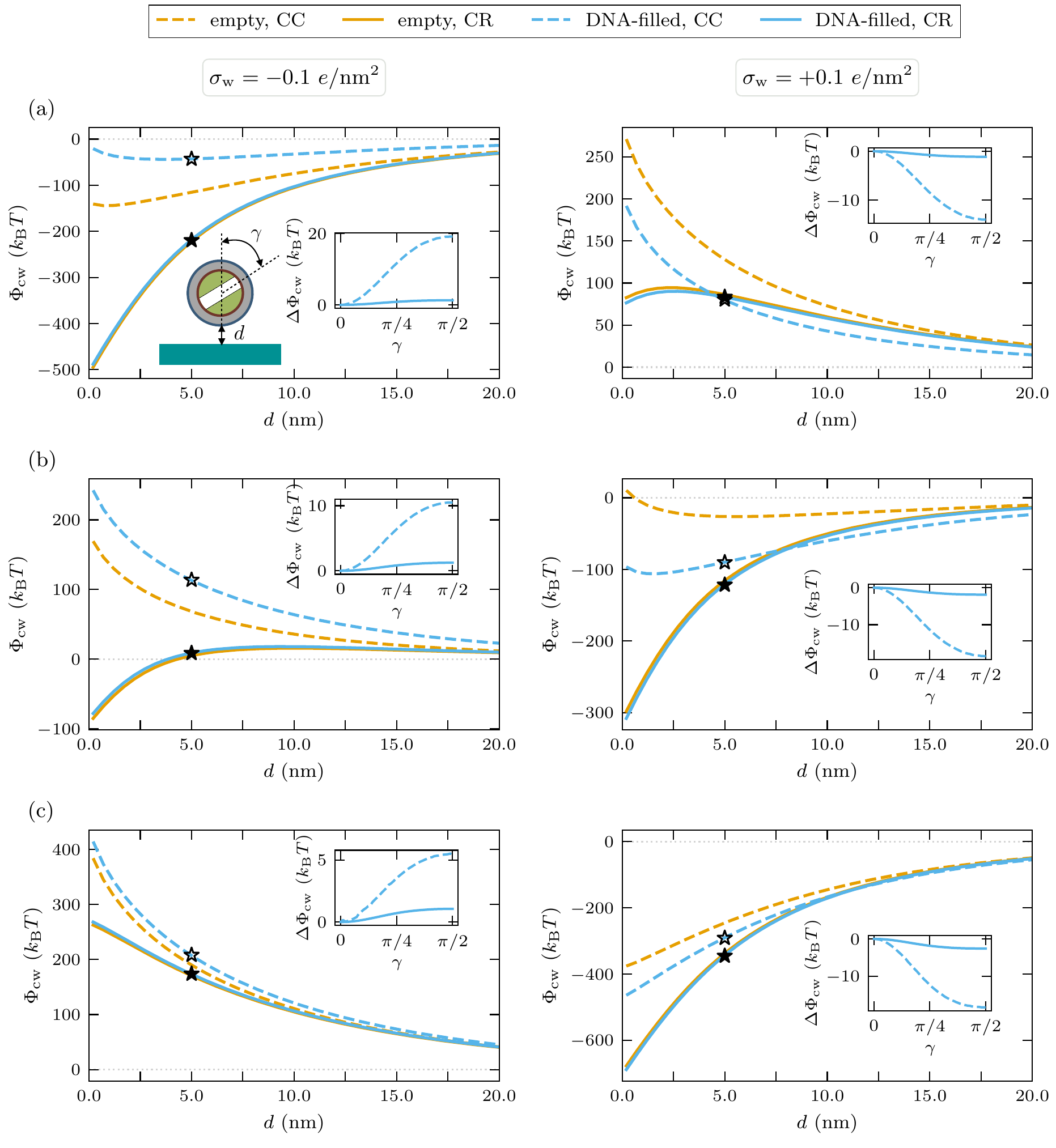}
\caption{Distance dependent virus--wall interaction potential $\Phi_\mathrm{cw}$ of empty and DNA-filled capsids. The interaction is shown at $\kappa^{-1}=10$~nm and for negative (left column) and positive (right column) surface charge density at (a) $\mathrm{pH=4}$, (b) $\mathrm{pH=5}$, and (c) $\mathrm{pH=6}$. Both CC and CR surface boundary conditions are considered. The DNA spool is aligned perpendicularly to the charged wall ($\gamma=0$). Insets in each panel show the angular dependence of the interaction potential of a DNA-filled capsid with the wall at $d=5$~nm in the form of $\Delta \Phi_\mathrm{cw}(\gamma)=\Phi_\mathrm{cw}(\gamma)-\Phi_\mathrm{cw}(\gamma=0)$, where the values of $\Phi_\mathrm{cw}(\gamma=0)$ are denoted in the panels with empty (CC boundary conditions) and full (CR boundary conditions) star symbols.
}
\label{fig:5}
\end{figure*}

\subsection{Interaction between a virus and a charged wall}
\label{sec:cw}

We first consider the geometry consisting of a charged wall with relative permittivity $\varepsilon_\mathrm{w}=2$ located at the half-space $z<0$, with the virus located at position ${\bf R}$ with the spool axis orientation $\hat{\bf u}$ (inset of Fig \ref{fig:5}(a)).  The wall carries fixed uniform surface charge that can either have positive or negative sign; specifically, we choose $\sigma_\mathrm{w}=\pm 0.1\,e/\mathrm{nm}^2$. The free energy of the system is
\begin{gather}
e^{-\beta F_\mathrm{cw}}\propto \int \mathrm{d}{\bf R}\int \mathrm{d}{\bf \hat{u}}\, e^{-\beta v(d)}\int\mathcal{D}\psi\, e^{-\beta\mathcal{F}_\mathrm{cw}[\psi]}\nonumber \\
=: \int \mathrm{d}{\bf R}\int \mathrm{d}{\bf \hat{u}}\, e^{-\beta H_\mathrm{cw}({\bf R},\hat{\bf u})}, \label{eq:effwall}
\end{gather}
which defines the effective Hamiltonian $H_\mathrm{cw}$ between the wall and the virus. Furthermore, $v(d)$ is the bare hard-core interaction potential, equal to $v(d)=\infty$ for $d<0$ and zero otherwise, where $d$ is the surface-to-surface separation.  The electrostatic free energy functional is identical to the single-capsid functional $\mathcal{F}_\mathrm{c}[\psi]$ of Eq.~\eqref{eq:freeenergy} with the exception that (i) the capsid geometry is taken at center-of-mass position ${\bf R}$ with orientation $\hat{\bf u}$; (ii) $\mathcal{V}_4$ is confined to the half-space $z>0$; and (iii) a wall contribution is added, given by
\begin{align}
\mathcal{F}_{\mathrm{w}}[\psi]&=
- \frac{1}{2} \varepsilon_0  \varepsilon_\mathrm{w}\int_{z<0} \dd V\,   |\nabla \psi({\bf r})|^2+\int_{z=0} \dd S\, \sigma_\mathrm{w}\psi({\bf r}).
\label{eq:wall}
\end{align} 
Furthermore, since the direction of the spool axis can now vary, the director in $\mathcal{V}_2$ is given by
\begin{equation}
\hat{\bf n}({\bf r})=\frac{\hat{\bf u}\times({\bf r}-{\bf R})}{|\hat{\bf u}\times({\bf r}-{\bf R})|}.
\end{equation}
Due to the symmetry of the system we parametrize the problem in terms of $d$ and angle $\gamma=\mathrm{arccos}(\hat{\bf u}\cdot\hat{\bf z})$. We can then infer an effective pair potential $\Phi_\mathrm{cw}$ by subtracting the self-energy contributions as $\Phi_\mathrm{cw}(d,\gamma)=H_\mathrm{cw}(d,\gamma)-H_\mathrm{cw}(d\rightarrow\infty)$, where $H_\mathrm{cw}$ is given within the mean-field approximation by
\begin{gather}
H_\mathrm{cw}(d,\gamma)=v(d)+\min_{\psi}\mathcal{F}_\mathrm{cw}[\psi;d,\gamma].
\end{gather}
A closed-form expression can be found for $\Phi_\mathrm{cw}$ in terms of $\phi({\bf r})$ which is determined by solving the underlying EL equations in this geometry for a given $d$ and $\gamma$. We refer to the SM~\cite{SM} for the analytical expressions and details of numerically determining $\Phi_\mathrm{cw}$.

To map out the behavior of $\Phi_\mathrm{cw}(d,\gamma)$ we first focus on the axisymmetric geometry where the capsid's spool axis is fixed to be parallel to the wall's surface normal $(\gamma=0)$. We start by discussing the case where the virus capsid is charge regulating. By varying $d$ we find for large separations---where the double layers weakly overlap, $\kappa d\gg 1$---that $\Phi_\mathrm{cw}$ is attractive or repulsive depending on the sign of the fixed wall charge $\mathrm{\sigma}_\mathrm{w}$ and on the total charge of the capsid (which includes DNA charge, free ions, and charges from amino acids). For example, for $\mathrm{pH}=4$ we find that both empty and filled capsids are globally positively charged which results in attraction (repulsion) for large enough $d$ when $\sigma_\mathrm{w}<0$ ($\sigma_\mathrm{w}>0$), see the full lines in Fig.~\ref{fig:5} at large $d$.  For some values of pH, the same behavior of $\Phi_\mathrm{cw}$ extends also to small $d$ where the electric double layers start to overlap. However, close to the isoelectric point of the outer shell (for $\kappa^{-1}=10$ nm we find $\mathrm{pI}\approx 4.7$, Fig.~\ref{fig:2}(a)), we find that the situation is reversed: in Fig.~\ref{fig:5}(a) (right panel) and Fig.~\ref{fig:5}(b) (left panel) the interaction is attractive at small $d$ and becomes repulsive at large $d$. For this to happen, a patch of opposite sign has to be induced on the outer shell which for a negative charged wall occurs for $\mathrm{pH}>\mathrm{pI}$ and for a positive wall at $\mathrm{pH}<\mathrm{pI}$. This is not an inference: Figs.~S7 and S8 in SM~\cite{SM} show the surface charge of the capsid facing the wall reversing sign directly, for pH above the isoelectric point with a negative wall and below it with a positive one. Comparison with the corresponding empty capsids (Figs.~S9 and S10 in SM~\cite{SM}) shows that this reversal is a response of the regulating shell to the wall rather than an effect of the genome. We note that the local reversal of charge can be inferred from applying Le Chatelier's principle on Eq.~\eqref{eq:reactions}; similar effects were reported in Refs.~\cite{Everts:2016}. Furthermore, we analyse the multipoles of the induced surface potentials in the SM \cite{SM} for both empty and filled capsids. The latter analysis shows that the induced charge density generally allows for odd multipoles, which are forbidden in the single-capsid case by symmetry. Furthermore, for the particular orientation we have chosen ($\gamma=0$), we find little variation between empty and filled capsids on the scale of the total interaction potential---the blue and orange full lines are nearly identical.

As a point of contrast, we also examine the CC boundary condition, where we use, at a fixed pH and $\kappa^{-1}$, the same charge as the one of an isolated empty capsid at $d\rightarrow\infty$. Contrary to the CR case, we find for $\gamma=0$ that filled capsids interact with the wall in a markedly different fashion from empty capsids, shown by the dashed blue and orange lines in Fig.~\ref{fig:5}. This difference becomes smaller when the surface charges on the capsid become large, where the CC case behaves similarly as the CR case, see Fig.~\ref{fig:5}(c).

Next, we assess how much of the anisotropy of a DNA-filled capsid translates into an observable torque on the capsid caused by the wall---which is one of the main findings of this work. We fix $d=5$ nm (i.e. $\kappa d=0.5$) where there is a strong double layer overlap; the distance is denoted by the star-shaped symbols in Fig.~\ref{fig:5}. Using fully three-dimensional finite-element calculations, we then vary $\gamma$, where $\Phi_\mathrm{cw}(d,\gamma)=\Phi_\mathrm{cw}(d,\pi-\gamma)$ due to up--down symmetry of our system. From the resulting angle-dependent interaction potential (insets in Fig.~\ref{fig:5}), we see that the anisotropy in $\Phi_\mathrm{cw}$ can amount to a few $k_\mathrm{B}T$ for the CR case, while it is more pronounced for the CC case (up to $\sim 10\, k_\mathrm{B}T$). CR therefore masks part of the anisotropy: a surface that is free to retitrate can absorb some of the interior inhomogeneity that a fixed-charge surface must pass on to the surrounding electrolyte. This is the same effect seen in Sec.~\ref{sec:singlecapsid} at the level of a single particle, where CR reduced the angular variation of $\phi_\mathrm{out}$ relative to the fixed-charge case, and it means the CC results should be read as an upper bound rather than as an alternative physical scenario.

Two features of this result are worth emphasizing. The preferred orientation is determined by the sign of the wall charge alone: it is $\gamma=0$ for a negative wall and $\gamma=\pi/2$ for a positive one at every pH we examined, even though the net charge of the capsid itself reverses between $\mathrm{pH}=4$ and $\mathrm{pH}=5$. And the effect is absent for an empty capsid, whose interaction with the wall has no orientational dependence at all. An orientational bias of a few $k_\mathrm{B}T$ is not large, but it is the kind of quantity that governs how particles deposit on charged substrates~\cite{Armanious2016}, and it is a property that only a filled capsid possesses. Since the CR case describes the physical system of interest, we checked whether the anisotropy can be increased by tuning $\sigma_\mathrm{w}$; we find that $\Phi_\mathrm{cw}$ becomes more repulsive or attractive without a significant change in its orientational part.

\subsection{Virus--virus interaction landscape}
\label{sec:pc}
Next, we consider a pair of identical viruses (labelled by A and B) with center-of-mass positions ${\bf R}_\mathrm{A}$ and ${\bf R}_\mathrm{B}$ and spool-axis orientations ${\bf \hat{u}}_\mathrm{A}$ and ${\bf \hat{u}}_\mathrm{B}$. Analogous to the virus--wall case described previously, the effective pair potential $\Phi_\mathrm{cc}$ is given by $\Phi_\mathrm{cc}(d,\hat{\bf u}_\mathrm{A},{\bf \hat{u}}_\mathrm{B})=H_\mathrm{cc}({\bf R}_\mathrm{AB},\hat{\bf u}_\mathrm{A},{\bf \hat{u}}_\mathrm{B})-H_\mathrm{cc}(d\rightarrow\infty)$, where ${\bf R}_\mathrm{AB}={\bf R}_\mathrm{A}-{\bf R}_\mathrm{B}$ and $d=|{\bf R}_\mathrm{AB}|-2R_\mathrm{out}$ is again the surface-to-surface separation. Within the mean-field approximation, we have
\begin{gather}
H_\mathrm{cc}({\bf R}_\mathrm{AB},\hat{\bf u}_\mathrm{A},{\bf \hat{u}}_\mathrm{B})=v(d)+\min_{\psi}\mathcal{F}_\mathrm{cc}[\psi;{\bf R}_\mathrm{AB},\hat{\bf u}_\mathrm{A},{\bf \hat{u}}_\mathrm{B}].
\end{gather}
A closed-form analytical expression for $H_\mathrm{cc}$ is presented in the SM~\cite{SM}, including details of its numerical evaluation. The free energy functional of two viruses $\mathcal{F}_\mathrm{cc}$ can be constructed from two single virus ones (see Sec.~\ref{sec:singlecapsid}) where one capsid is positioned at a fixed ${\bf R}_\mathrm{A}$ with orientation $\hat{\bf u}_\mathrm{A}$ and the other one at ${\bf R}_\mathrm{B}$ with $\hat{\bf u}_\mathrm{B}$. Outside the two viruses is again a region filled with water and free ions.

\begin{figure}
\centering
\includegraphics[width=0.47\textwidth]{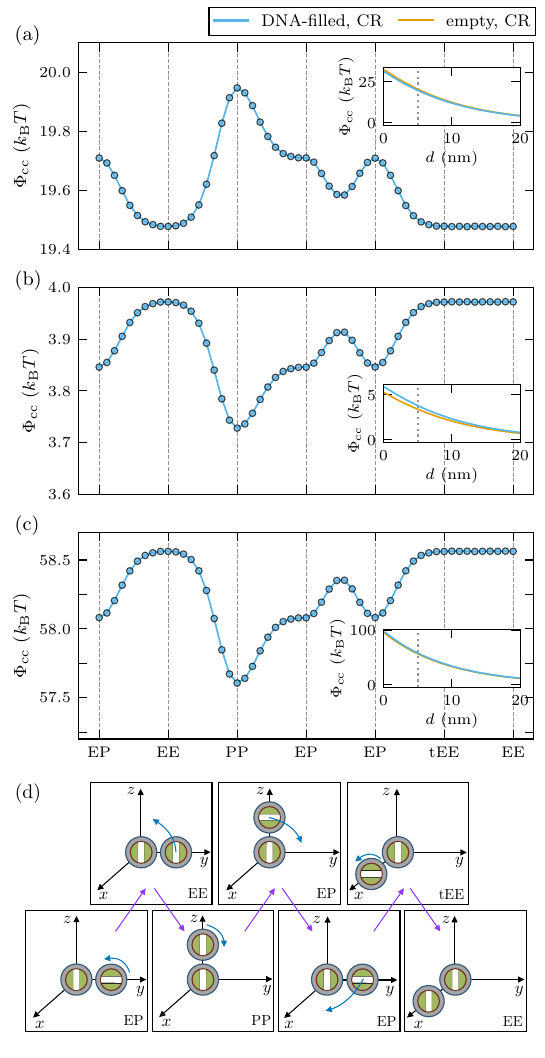}
\caption{Orientation-dependent virus--virus interaction potential $\Phi_\mathrm{cc}$ at fixed surface-to-surface separation $d=5$ nm for two charge-regulating DNA-filled capsids at (a) $\mathrm{pH}=4$, (b) $\mathrm{pH}=5$, and (c) $\mathrm{pH}=6$. Schemes of the various orientations and what rotations are used to transition between them are shown in panel (d) and correspond to the gray dashed lines in panels (a)--(c). The three distinct configurations are labelled by equator (E), pole (P), and transverse (t). For the PP orientation, we show in the insets of panels (a)--(c) how $\Phi_\mathrm{cc}$ varies with $d$ and compare them with an empty capsid for which $\Phi_\mathrm{cc}$ does not depend on orientation.}
\label{fig:6}
\end{figure}

First, we consider the axisymmetric setting as a function of $d$ where both capsids are oriented with their spool and separation axes parallel to each other---the so-called pole--pole (PP) orientation, $\hat{\bf u}_\mathrm{A}\parallel\hat{\bf u}_\mathrm{B}\parallel{\bf R}_\mathrm{AB}$. For simplicity, we focus purely on identical, charge regulating viruses as they are the most biologically relevant. As a function of $d$ (see insets of Fig.~\ref{fig:6}), we observe an expected purely repulsive potential, which for large $d$ behaves as a screened Coulomb interaction between effective (point) charges. We also see that the DNA-filled case significantly differs from the empty-capsid case only close to the isoelectric point (Fig.~\ref{fig:6}(b)) in this particular virus--virus orientation. Furthermore, for a filled capsid we have a modulation with the familiar anisotropy function, see Refs.~\cite{Trizac:2000,Tellez:2010}. However, this function captures the anisotropy at large separations, whereas in this work we are interested in configurations where the electric double layers overlap: for such configurations, the anisotropy is stronger and is therefore expected to be directly measurable.

The interesting part of the virus--virus geometry is that the configurational space for interactions is much larger than in the virus--wall case. To map out the orientational dependence for DNA-filled capsids, we fix $\kappa d=0.5$, and use the visualisation proposed in Ref.~\cite{Gnidovec:2025}. Here, one virus is fixed at the origin with spool axis in the $z$ direction while the second virus rotates either by changing its separation vector ${\bf R}_\mathrm{AB}$ or by changing its spool axis, see Fig.~\ref{fig:6}(d). The one-dimensional projections of $\Phi_\mathrm{cc}$ obtained in this manner are plotted in Fig.~\ref{fig:6}(a)--(c) for three distinct pH values. We observe that the difference between global maximum and minimum in these cases can go up to 1~$k_\mathrm{B}T$, which is generally lower than in the particle-asymmetric virus--wall geometry (Fig.~\ref{fig:5}, insets). Nevertheless it provides a potentially measurable effect. Furthermore, when the isoelectric point for empty capsids is crossed by increasing the pH, we find that the slopes of the curves change sign, suggesting a reversal of torques that is tunable with pH---compare panel (a) with panels (b) and (c) in Fig.~\ref{fig:6}. Moreover, the equilibrium orientation at fixed $d$ changes from equator--equator (EE; Fig.~\ref{fig:6}(a)) to PP (Fig.~\ref{fig:6}(b),(c)). With a modulation below $k_\mathrm{B}T$, an isolated pair is biased towards these configurations rather than locked into them, and the switch is unlikely to be resolvable pair-by-pair against rotational Brownian motion. It should instead appear as a pH-dependent orientational correlation in suspensions dense enough for the bias to act between many neighbors, which is the setting we return to in Sec.~\ref{sec:exp}. It is worth noting where the orientational part of the interaction is largest relative to the interaction as a whole. The isotropic baseline varies by more than an order of magnitude between pH 4 and 6 while the modulation does not, so the ratio of the two is greatest near the isoelectric point, which is where any experiment to study this system should be designed at as well.

\section{Experimental implications}
\label{sec:exp}

The anisotropic exterior charge and potential we have described are, in principle, accessible to several established single-particle and ensemble techniques. We discuss three routes in what we regard as increasing order of feasibility---covering direct measurements, electrokinetic measurements, and interaction and phase behavior---and in each case we try to state the size of the predicted signal rather than solely its existence.

\subsection{Direct measurements}

The most direct route to study the effects described in this work would be to map the electrostatic potential around a single capsid. Such measurements are becoming possible: local electric fields have recently been visualized using fluorescent ions in a ferroelectric nematic liquid crystal~\cite{Mertelj:2026}, and interfacial potentials can be probed spectroscopically~\cite{Bonn:2025} or by iontronic microscopy~\cite{Faez:2026}. Scanning dielectric microscopy can already detect single viruses through their polarization response~\cite{Fumagalli2012}. The signal we predict, however, is demanding: under CR, the outer surface potential differs between the polar caps and the equatorial band by only a few percent of $k_\mathrm{B}T/e$, over caps a few tens of nanometres across (Fig.~\ref{fig:4}(b)). We therefore expect the first evidence to come not from the potential itself but from quantities derived from it, which is what we describe next.

\subsection{Electrokinetic measurements} 

At the level of a single DNA-filled capsid, we found an inhomogeneous surface potential distribution on the outer capsid surface. The question is whether these effects are measurable using external electric fields. For a particle of \emph{any} shape with a uniform surface or zeta potential $\psi_0$, it is well known that the translational velocity ${\bf U}$ and rotational velocity $\boldsymbol{\Omega}$ in a uniform external electric field ${\bf E}$ are given by the Smoluchowski formulas~\cite{Smoluchowski} (see Ref.~\cite{Cichockibook} for an English translation) ${\bf U}=(\varepsilon\psi_0/\eta){\bf E}$ and $\boldsymbol{\Omega}=\boldsymbol{0}$ for a thin electric double layer. Here, $\eta$ is the dynamical shear viscosity of the solvent (water). Teubner~\cite{Teubner:1982} generalised these relations to an inhomogeneous surface potential. Applying his result to a sphere of radius $R_\mathrm{out}$ with an arbitrary axisymmetric surface potential $\psi(\vartheta)=\sum_lc_lP_l(\cos\vartheta)$ in an unbounded fluid, and in the thin-double-layer limit $\kappa R_\mathrm{out}\gg1$, one finds a selection rule. The angular weights entering the translational velocity are quadratic in $\cos\vartheta$, so $\mathbf U$ depends only on $c_0$ and $c_2$; the weight entering the rotational velocity is linear, so $\boldsymbol\Omega$ depends only on $c_1$. Three consequences follow. The single-capsid spool has no odd multipoles, so $\boldsymbol\Omega=\mathbf 0$ and there is no electrorotation in a uniform field. The mobility acquires an orientational dependence of relative size $\propto c_2/c_0$, which for the values in Fig.~\ref{fig:4} amounts to about a percent---measurable in principle, but close to the limit of what is accessible. And, importantly for interpretation, the orientational average of $\mathbf U$ is exactly the Smoluchowski result with $c_0$ alone: ensemble mobility measurements are blind to the imprint and report only the monopole, so that orientation-resolved or single-particle measurements are required. This also means that the difference between empty and filled capsids reported in bulk electrophoresis~\cite{heffron2021virus,Michen2010} is a statement about net charge, not about the anisotropy discussed here.

In a non-uniform field the selection rule no longer applies, since quadrupoles can experience torques~\cite{Jones:1996}. Dielectrophoretic alignment would therefore give access to $c_2$ directly, and would distinguish a filled capsid from an empty one whose exterior field has no quadrupole at all. This seems to us the most promising single-particle route, and we note that single-particle charge metrology of empty and filled capsids is already established in a different setting~\cite{Jarrold2022,heldt2023empty}.

\subsection{Interaction and phase behavior} 

The orientation-dependent virus--virus pair potential of Sec.~\ref{sec:multicapsids} enters the structure factor of concentrated virus suspensions, and at sufficient density their phase behavior. This is the setting in which a sub-$k_\mathrm{B}T$ bias is most likely to become visible, because it acts between every pair of neighbors rather than only once. Virus suspensions are well established as model systems for exactly this kind of measurement~\cite{Fraden1995}, and small-angle X-ray scattering and osmotic-stress measurements on ordered arrays give access to the relevant correlations: the anisotropic component of $\Phi_\mathrm{cc}$ contributes an orientational correlation that is absent for an empty capsid. Direct force measurements between tethered particles by AFM or optical tweezers offer a single-pair alternative. As we discussed in Sec. \ref{sec:cw}, the orientational modulation on the interaction potential is largest when a constant-charge probe is used, as charge regulation would mask the generation of torques. Furthermore, we note that the interaction potential can also be probed by tracking the Brownian motion of a pair of particles or a single particle next to a charged wall, as was done in Ref. \cite{everts2021anisotropic}.
 
In these settings, the discriminating observable can be made a difference rather than an absolute value. The cleanest discriminants are the modulation of the signal with pH and with genome content, both experimentally controllable and both predicted quantitatively by the model---as well as both absent for an empty capsid. Comparing preparations that differ only in packaged genome length---routinely separated in gene-therapy vector production~\cite{heldt2023empty,Jarrold2022}---would isolate the genome contribution directly.

\section{Discussion and conclusion}
\label{sec:dis}

We have studied whether a packaged genome can imprint an electrostatic anisotropy to a virus that its capsid cannot on its own. In the latter case, an empty icosahedral capsid possesses no anisotropic multipole below $l=6$ by virtue of symmetry, and screening removes such high multipoles at large distances. A genome is not bound by that symmetry, and we have shown that a dsDNA genome packed as an inverse spool supplies an anisotropic footprint of its geometry. What the virus presents to the environment is essentially the electrostatic ``shadow'' of the DNA-free void at the center of the spool, blurred over an angular width comparable to the shell thickness seen from the capsid center. The angular shape of the imprint is therefore fixed by the packing geometry, and the electrostatic state of the system---pH, ionic strength, the charge and dielectric response of the DNA---enters only through the amplitude. Electrostatic screening then acts as a filter on that shape: the quadrupolar and hexadecapolar components are comparable on the capsid surface, but they decay at very different rates. Beyond a few screening lengths the multipole ratios saturate at values set by $\xi_l(\kappa R_\mathrm{out})$ alone, leaving a predominantly quadrupolar far-field. Salt thus controls not only how far the field reaches, but which part of its angular structure is retained. The quadrupole that does survive is enough to make the interaction of a filled capsid with a charged wall, and with another capsid, depend on orientation at the $k_\mathrm{B}T$ level, with a preferred orientation set by the sign of the external charge and no counterpart for an empty capsid.

Our results relate directly to the measured differences between empty and filled capsids that motivated the work~\cite{hernando2015quantitative,heldt2023empty,caniglia2022probing}. The shift in the mean surface charge and potential upon filling does not require the genome to be ordered: any interior charge raises the magnitude of the potential throughout the particle, and a charge regulating shell responds by retitrating. That part of the puzzle is resolved by CR alone and requires no transfer of charge across the shell. What is specific to an ordered genome is the anisotropy, which those measurements did not resolve and which is the prediction we provide here. The distinction matters in practice and cannot be tested by a measurement of net charge but requires a measurement of an orientation-resolved observable.

Several limitations bound our findings and provide natural extensions of the work. The treatment is mean-field, with monovalent salt; the ion–ion correlations and multivalent effects that drive DNA condensation in many real systems are outside its scope. We assume a single, fully developed nematic configuration, whereas packaged DNA samples a family of conformations~\cite{coshic2024structure,leforestier2013polymorphism,liu2014solid,petrov2007conformation,farrell2024spool}. The degree of order enters the in-plane permittivity of the composite, whereas the angular shape of the imprint follows from the architecture of the packing, so imperfect ordering is expected to change the amplitude of the imprint rather than its geometry. Related to this, cryo-EM and scattering studies indicate that even at full packing the genome is not uniformly ordered: an outer, hexagonally ordered shell surrounds a core of less ordered DNA at lower
density~\cite{lander2013dna,leforestier2013polymorphism,Leforestier2010}, and recent work suggests that twist rather than bend dominates its liquid-crystalline organization~\cite{Twist2025}. Treating the core as DNA-free is an idealization that affects the contrast, and hence the amplitude of the
imprint, more than its angular structure. For the same reason we restrict physical conclusions to the filling range in which the packaged DNA is expected to be ordered.

Furthermore, we note that the densely packaged DNA within the virus is highly charged: future improvements of the model should therefore account for the finite ion size, which prevents electrostatic potentials from becoming too large as is common in PB type theories. This can be done, for example, by a straightforward Bikerman-type modification of the free energy \cite{Bikerman:1942,Orland:1997}. A second parameter deserving scrutiny is the dielectric response of the DNA itself. We take $\varepsilon^c=2$, whereas direct nanoscale measurements on isolated DNA give a somewhat higher value~\cite{Cuervo2014}; the permittivity of water confined between closely packed helices is in turn lower than in bulk~\cite{Fumagalli2018}. These uncertainties bound the dielectric contrast of the composite and therefore the amplitude of the imprint, but they leave $\vartheta_c$, and with it the angular structure, untouched. In addition, extending the analysis to the interplay between the DNA-induced quadrupole and a genuinely heterogeneous surface-charge distribution is a clear next step; the ionizable residues are here distributed uniformly over each surface, and the analysis can be extended to the icosahedrally patterned distribution of a real capsid and to its interplay with the genome-induced quadrupole.

Finally, because the director field differs from genome to genome~\cite{coshic2024structure,petrov2007conformation}, the specific form of the exterior imprint may be virus-specific, which suggests an electrostatic route to distinguishing packaged-genome architectures without resolving them structurally. More broadly, nothing in the mechanism is particular to viruses. Any
orientationally ordered charged medium confined by a dielectric shell will transmit the symmetry of its ordering, rather than its charge, to the exterior, and an electrolyte will filter that symmetry down to its lowest surviving multipole. Viruses are the case in which the confinement is tightest and the ordering best characterized, but the same reasoning applies to the deliberate loading of virus-like particles with ordered or
patchy cargo~\cite{Chen2006,Ting2011,Muhren2023}, and to archaeal dsDNA viruses~\cite{zhang2025cryo} whose packing architectures are only now being resolved.

\section*{Acknowledgements}
We regret to inform the reader that our co-author and conceiver of this project, Professor Rudolf Podgornik, passed away on 28 December 2024. J.~C.~E.\ acknowledges funding from the National Science center, Poland, within SONATA BIS grant no.\ 2023/50/E/ST3/00452. A.~B.\  acknowledges funding from the Slovene Research and Innovation Agency (ARIS) under contracts no.\ P1-0055 and no.\ J1-60002.

\section*{Author contributions}

R.~P.\ conceived the project and created the first version of the model. Derivations were performed by R.~P.\ and J.~C.~E. The data curation was performed by A.~B.\ to choose the model parameters; A.~B.\ was also responsible for the visualisations. J.~C.~E.\ performed the numerical computations. A.~B.\ and J.~C.~E.\ performed the formal analysis, validation of the model, and were responsible for writing the first draft. Both of them reviewed and edited the final manuscript.

\section*{Data availability}
Details of the numerical procedures, COMSOL settings, extended derivations, and additional figures and data are presented in the Supplemental Material~\cite{SM}.

\bibliography{bibliography}

\end{document}